\documentclass[twocolumn, english]{article}
\usepackage[T1]{fontenc}
\usepackage[latin9]{inputenc}
\usepackage{float}
\usepackage{amsmath}
\usepackage{graphicx}
\usepackage{amssymb}
\usepackage{epstopdf}
\usepackage{subfigure}
\makeatletter

\usepackage{float}\usepackage{babel}
 \usepackage{subfigure}

\makeatother

\usepackage{babel}

\newcommand\CGRplot[1]{%
      \begin{tabular}{@{}c@{}c@{}c@{}}
          C && G \\[-2pt]
          &\includegraphics[width=0.262\textwidth]{#1}& \\[-3pt]
          A && T
      \end{tabular}}

\begin{document}

\title{Map of Life:  Measuring and Visualizing  Species' Relatedness with  {\it Molecular Distance Maps}}

\author{Lila Kari$^{1, 2}$
\and Kathleen A. Hill$^3$
\and Abu Sadat Sayem$^2$ 
\and Nathaniel Bryans$^4$
 \and Katelyn Davis$^3$
\and Nikesh S. Dattani$^5$}

\maketitle

\footnotetext[1]{Corresponding author, email lila.kari@uwo.ca}
\footnotetext[2]{Department of Computer Science, 
University of Western Ontario, London, ON, N6A 5B7 Canada}
\footnotetext[3]{Department of Biology, University of Western Ontario, London, ON, N6A 5B7 Canada} 
\footnotetext[4]{Microsoft Corporation, Redmond, WA,  98052, USA}
\footnotetext[5]{Physical and Theoretical Chemistry Laboratory, Dept. of Chemistry, Oxford University,
 Oxford, OX1 3QZ, UK}

\begin{abstract}

We propose a  novel combination of methods that {\it (i)} portrays quantitative characteristics of a DNA
 sequence  as an image,  {\it (ii)} computes distances between
  these images, and {\it (iii)}  uses these distances to output a map wherein each sequence is
 a point in a common Euclidean space. In the resulting   {\em Molecular Distance Map} each
 point signifies a DNA sequence,  and the  geometric distance between any two points reflects the degree
 of relatedness between the corresponding sequences and species.

{\it Molecular Distance Maps}  present compelling visual  representations of relationships
 between species  and could be used  for taxonomic clarifications, for species identification,
 placement of species in existing taxonomic categories, as well as for studies of evolutionary 
history. One of the advantages of this method is its general applicability since, as 
sequence alignment is not required, the  DNA  sequences  chosen for comparison can be completely
different regions in different genomes.  In fact, this method can be used to compare {\it any} 
two DNA sequences. For example,  in our dataset of 3,176  mitochondrial DNA sequences, 
 it correctly finds the mtDNA sequences most closely related  to that of the anatomically modern
 human (the Neanderthal, the Denisovan, and the chimp), and it finds that the sequence most different
 from it  belongs to a cucumber. Furthermore, our method can be used to compare real sequences to artificial, computer-generated, DNA sequences. For example, it is used to determine that the distances between  a {\it Homo sapiens sapiens} mtDNA and
artificial sequences of the same length and same trinucleotide frequencies can be larger than the distance
between the same human mtDNA and the mtDNA of a fruit-fly.

We demonstrate this method's promising potential for taxonomical clarifications by applying
 it to a diverse variety of cases that have been historically controversial, such as the genus
 {\it Polypterus}, the family Tarsiidae, and the vast (super)kingdom  Protista.

\end{abstract}

\section{Introduction}
 In 2012 alone,  biologists  described  between 16,000 and 20,000 new species \cite{ScienceNews12}. 
Recent findings, \cite{Mora2011}, suggest that as many as 86\% of existing species on Earth
 and 91\% of species in the oceans have  not yet been  classified and catalogued. 
It is thus imperative to find a comprehensive,  quantitative,  general-purpose method to reliably
 identify
 the relationships among the 1.2 million species that have already been  catalogued,  
\cite{Mora2011}, as well as 
 the vastly larger numbers of those that have not. 

We propose a combination of three  techniques to efficiently measure distances between DNA
 sequences, and to simultaneously  visualize the  relationships among all the DNA sequences
 within any given dataset.
The result of applying this method to a  collection of DNA sequences is  a 
{\it Molecular Distance Map}  that  allows the visualization of the sequences as points
 in a common two-dimensional Euclidean space, wherein the geometric distance between any
  two points reflects the differences in composition between all the subsequences of the two
 sequences.
 The proposed method is based on the {\it Chaos Game Representation} (CGR) of DNA sequences,
 \cite{Jeffrey1990,Jeffrey1992},  a genomic signature that  has a remarkable ability
 to differentiate between genetic sequences belonging to different species, see Figure \ref{fig:CGR}.
  Due to this characteristic, a  {\it Molecular Distance Map} of a  collection of genetic sequences  
 allows the inferrence of relationships between the corresponding species.

Concretely, to compute and visually display relationships between DNA sequences in a given set 
 $S = \{s_1, s_2, ..., s_n \}$ of $n$ DNA sequences, we propose the following  combination of
 three techniques:

\begin{itemize}
\item {\em Chaos Game Representation} (CGR) to deterministically represent each DNA sequence
 $s_i$, $1\leq i\leq n$, as  a two-dimensional black-and-white image denoted by $c_i$;
\item {\em Structural Dissimilarity Index} (DSSIM), an image-distance measure,
  to compute the distances $\Delta (i, j)$, $1\leq i, j \leq n$, between   pairs of
 CGR images $(c_i, c_j)$, and produce a  distance matrix;
\item {\em Multi-Dimensional Scaling} (MDS), applied to the distance matrix to produce
 a  map in the Euclidean space wherein each plotted point $p_i$ with coordinates $(x_i, y_i)$ 
represents the DNA sequence $s_i$ whose CGR image is $c_i$. The position of the point $p_i$ in
 the map, relative to all the other points $p_j$,  reflects the  distances between the DNA sequence
 $s_i$  and the  DNA sequences $s_j$ in the dataset.
\end{itemize}

The combination of CGR, DSSIM, and MDS applied to any given set of genetic sequences 
yields a  {\em Molecular Distance Map} which visually  illustrates the
 quantitative relationships and patterns of proximities among the given 
 genetic sequences and, accordingly, among the species they  represent.

Besides presenting compelling visual pictures of the relatedness between species as seen in,  e.g., Figure~\ref{vertebrates}, 
 there are several advantages of our proposed method  over other methods  in computational phylogenetics.

Advantages over alignment-based methods  include the fact that  our method  allows
 comparison between {\it any}  two  DNA sequences.  In particular, it
 allows comparisons within the  genome of an individual, across genomes within a
 single species, between genomes within a taxonomic
 category, and across taxa, while also allowing the use of completely different portions
 of the genomes  for comparison. In addition, sequence alignment analyses use only a limited aspect
 of the compared sequences, namely the 
identity of characters at  each position, whereas our  method is based on significantly more information by 
simultaneously  comparing all subsequences of the given 
 sequences. Lastly, in our method, computing distances between different sequences 
 is completely automatized and does not required manual intervention or calibration.

An advantage over phylogenetic trees \cite{PhylogeneticHandbook}  pertains to the fact
that, in a phylogenetic tree, adjacency of two species-representing leaves is not always 
meaningful since one can rotate
branches about the nodes of the tree. In contrast, in a Molecular Distance Map the  distance on the
 plane between any two species-representing  points  has a concrete and fixed meaning. Furthermore, our
 DSSIM distance matrices can be used,  \cite{Fitch1967}, to calculate phylogenetic trees in addition to
 Molecular Distance  Maps.

Advantages over DNA barcodes \cite{Hebert2003} and Klee diagrams
 \cite{Sirovich2010}  include the fact that  Molecular Distance Maps are 
 not alignment-based, and that they are applicable to cases where barcodes may have limited effectiveness: Plants
 and fungi for which different barcoding regions have to be
 used \cite{Kress2005},   \cite{Hollingsworth2009}, \cite{Schoch2011};  
protists where multiple loci are generally needed to distinguish between species \cite{Hoef2012}; 
prokaryotes  \cite{Unwin2003};  and artificial, computer-generated,  DNA sequences.

Finally, our proposed approach  addresses the need for a visual representation of species relatedness
 that is easily interpretable, as well as  capable of including a vast amount of
data rather than selective or regional datasets.

\section{Methods}

\begin{figure*}[ht]
        \centering
        \footnotesize
        \subfigure{%
         }\hspace{.5cm}
        \subfigure{%
\CGRplot{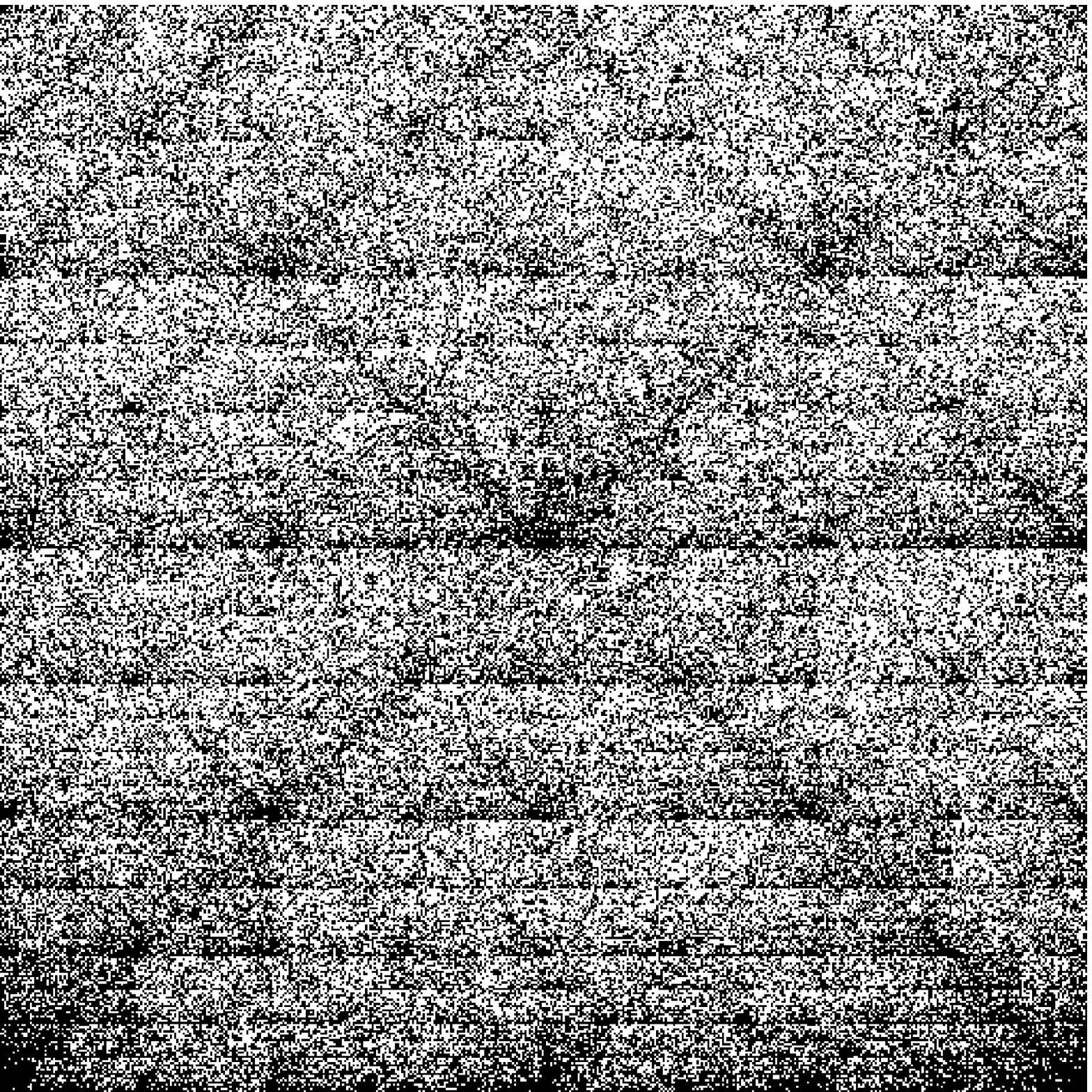}
        }\hspace{.5cm}
        \subfigure{%
\CGRplot{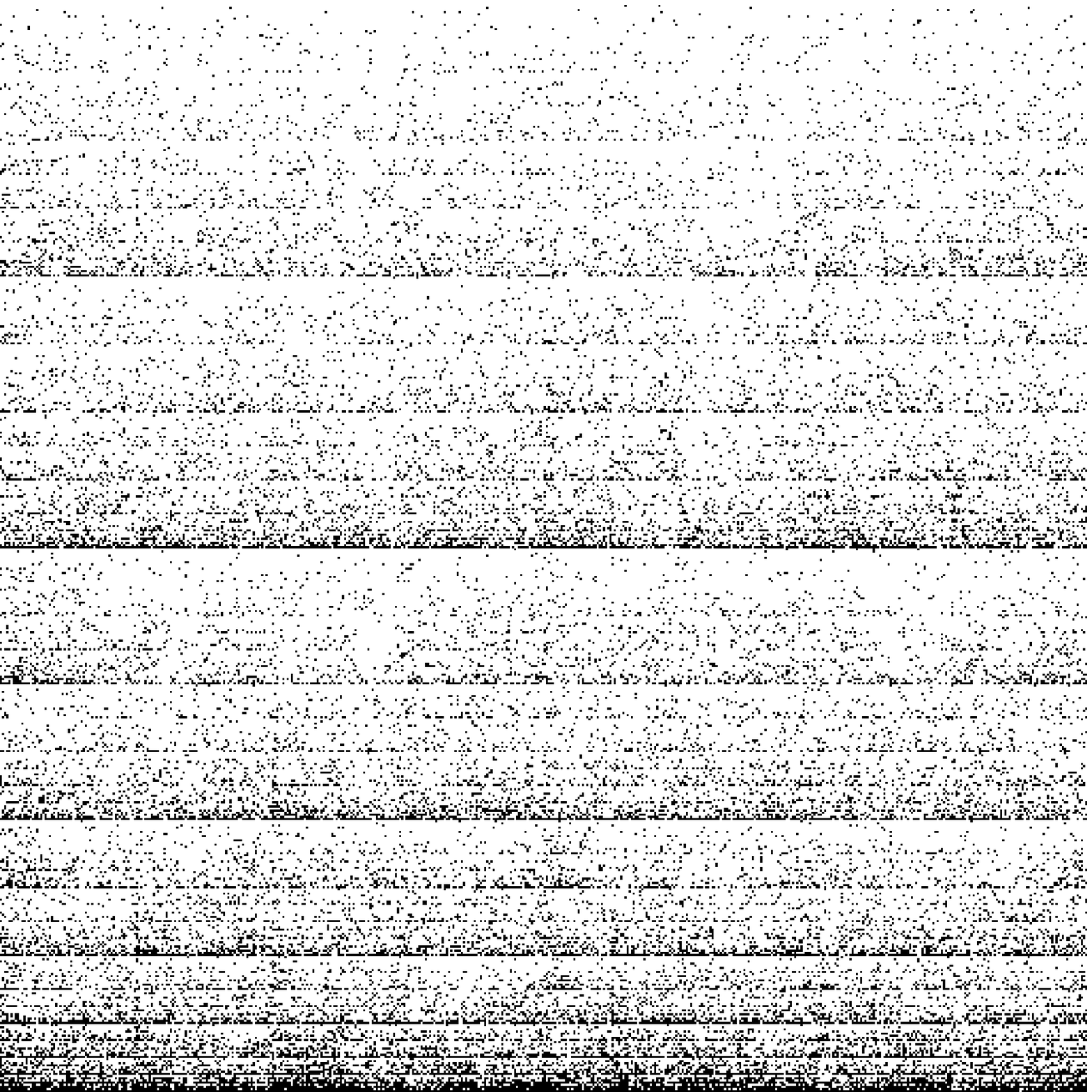}
        }\hspace{.5cm}
        \subfigure{%
\CGRplot{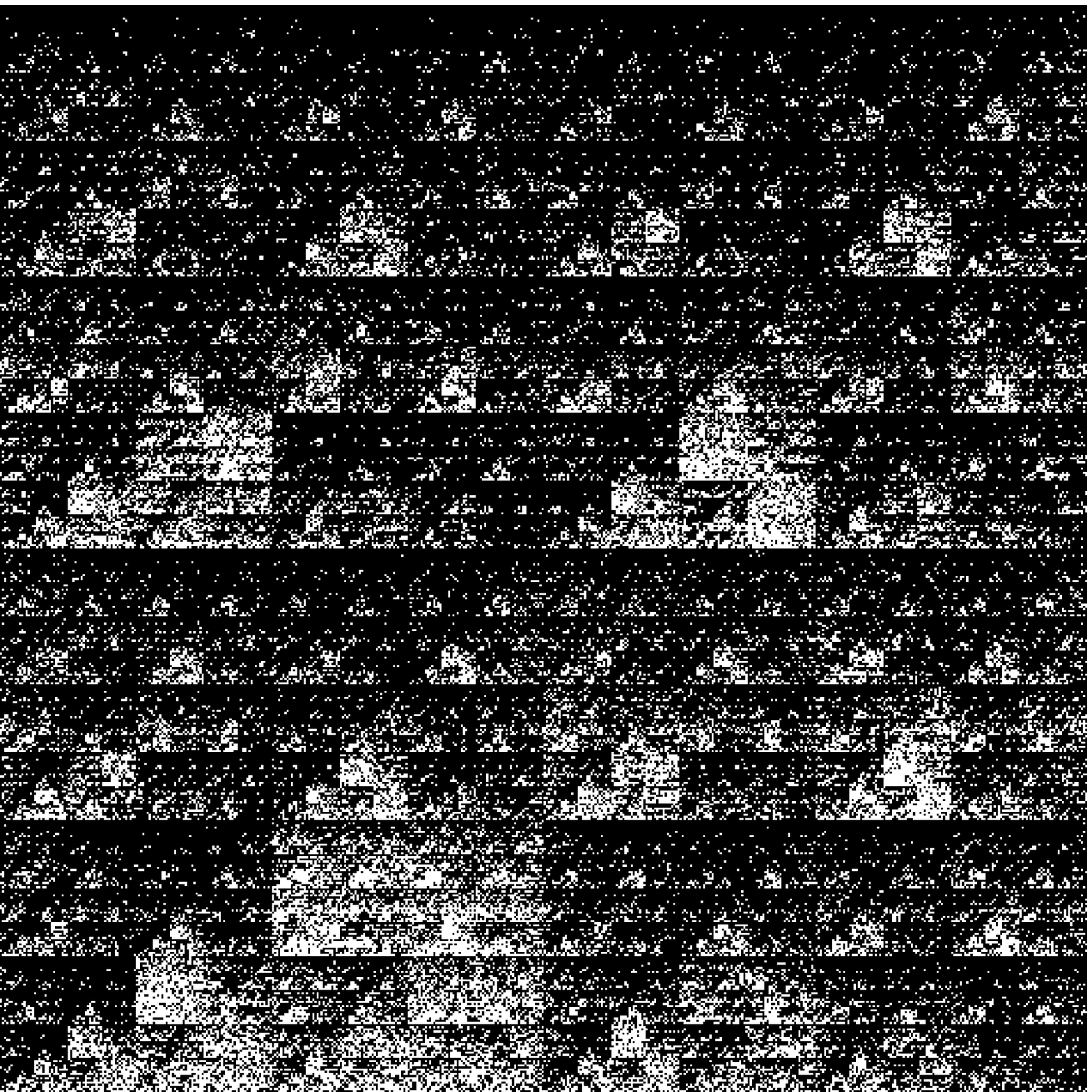}
        }
        \caption{CGR images for  various  genomes. From left to right: 
(1) {\it Marchantia polymorpha}  (liverwort) mtDNA, 186,609 bp;
(2)  {\it Malawimonas jakobiformis} (flagellate) mtDNA,  47,328 bp;
(3) {\it Rhodobacter capsulatus}, full genome, 3,738,958 bp.}
        \label{fig:CGR}
\end{figure*}

A CGR is a genomic signature that  utilizes all subsequence composition structures of
 DNA sequences, and is  genome and species specific, 
\cite{Jeffrey1990,Jeffrey1992,Hill1992,Hill1997,Deschavanne1999, Deschavanne2000,Wang2005}.
The sequences chosen from each genome as a basis for computing ``distances'' between
 species do not need to have any relation with one another from the point of view of their 
 position or information content. A CGR \cite{Jeffrey1990, Jeffrey1992}
 associates an image to each DNA sequence as follows: Starting from a unit square
 with vertices
 labelled {\it A, C, G,} and {\it T}, and the center of the square as the starting point, the image
 is obtained by  successively plotting each nucleotide as the middle point between the current point and the 
vertex labelled by the nucleotide to be plotted.
If the generated square has a resolution of $2^k\times 2^k$ pixels, then every pixel represents a distinct DNA
initial subsequence of length $k$; a pixel is black if the subsequence it represents occurs in the DNA sequence, 
otherwise it is white. In general 4,000 bp are necessary for a sharply  defined image, but in many cases 2,000 bp
give a reasonably good approximation, \cite{Jeffrey1990}.
CGR images of genetic DNA sequences originating from various species show interesting patterns such as squares,
 parallel lines, rectangles, triangles, and also complex fractal patterns, Figure~\ref{fig:CGR}.

Other visualizations of genetic data have been proposed, such as the 2D
 rectangular walk \cite{Gates1986}
 and  methods similar to it  in \cite{Nandy1994}, \cite{Leong1995},
  vector walk  \cite{Liao2005},  cell \cite{Yao2004},    vertical vector  
\cite{Yu2010},  Huffman coding  \cite{Qi2011}, and  colorsquare \cite{Zhang2012} methods.
 Three-dimensional representations of DNA sequences include  the tetrahedron   \cite{Randic2000}, 
 3D-vector  \cite{Yuan2003}, and  trinucleotide curve \cite{Yu2009}  methods.
 Among  these visualization methods, CGR images arguably provide the most immediately 
comprehensible  ``signature'' of a DNA sequence and a desirable genome-specificity, \cite{Jeffrey1990,Deschavanne2000}.
 In  addition, the  DNA representing black-and-white images produced by the  CGR method
  are easy to compare, both visually and computationally.

{\em Structural Similarity} (SSIM) index is an image similarity index  used in
 the context of image processing  and computer vision to compare two black-and-white images from the
 point of view of their structural similarities \cite{Wang2004}.  SSIM combines three parameters -
 luminance distortion, contrast distortion, and linear correlation - and was designed to perform 
 similarly to the human visual system, which is highly adapted to extract  structural information.   
Originally, SSIM  was defined as a similarity measure $s(A,B)$ whose theoretical range
between two  images $A$ and $B$ is $[-1,1]$ where a high
value amounts to close relatedness. We use a related {\em DSSIM distance}
$\Delta(A,B) = 1 - s(A,B) \in [0,2]$, with the distance being 0 between two
identical images, 1 between e.g.\ a black image and a white image, and 2 if
the two images are negatively correlated; that is, $\Delta(A,B) = 2$ if and
only if every pixel of image $A$ has the inverted value of the corresponding
pixel in image $B$ while both images have the same luminance (brightness).
 For our particular dataset of genetic CGR images, all distances ranged between 0 and 1.

MDS  has been used for the visualization of data relatedness  based on distance matrices in various
 fields such as cognitive science, information science, psychometrics, marketing, ecology, social science, and  other areas of study \cite{BorgGroenen2010}. MDS takes as input a distance matrix  containing
 the pairwise distances between $n$ given  items and outputs a two-dimensional map
 wherein each item is represented by a point, and the geometric distances between points reflect the
 distances between the corresponding items in the distance matrix. 
Two notable examples of molecular biology studies that used MDS are \cite{Lessa1990} (where it was
 used for the analysis of geographic genetic distributions of some natural populations) and \cite{Hebert2003}
 (where it was used to provide a graphical summary  of the distances among  CO1 genes from  various species).

Classical MDS, which we use in this paper, receives as input an $n\times n$ distance 
matrix  $(\Delta(i, j))_{1\leq i, j \leq n}$ of the pairwise distances between any
 two items
 in the set.
The output of classical MDS consists of $n$ points in a $q$-dimensional
 space whose pairwise Euclidean distances are a linear function of the
 distances between the corresponding items in the input distance matrix.
More precisely, MDS will return $n$ points $p_{1},p_{2},\ldots,p_{n}\in \mathbb{R}^{q}$ such that
 $d(i, j)= ||p_{i}-p_{j}||\thickapprox f(\Delta(i,j))$ for all $i,j\in \{1, \ldots, n \}$
 where $d(i, j)$ is the Euclidean distance between  the points $p_i $ and $p_j$, and $f$ 
is a  function linear in $\Delta(i, j)$. Here, $q$ can be at most $n-1$ and the points are recovered from
 the eigenvalues and eigenvectors of the input 
$n\times n$ distance matrix. If we choose $q=2$ (respectively $q=3$), the result of classic
 MDS is  an approximation of the original $(n-1)$-dimensional space as a two- (respectively three-) dimensional map.

In this paper all Molecular Distance Maps  consist of coloured points, wherein each point represents an mtDNA sequence from the dataset. Each mtDNA sequence is
  assigned a unique numerical identifier retained in all analyses, e.g., \#1321 is the identifier for
 the {\it Homo sapiens sapiens} mitochondrial genome. The  colour assigned to a sequence-point  may however  vary from map to map,
 and it depends on  the taxon assigned to the point in a particular Molecular Distance Map.
 For consistency, all maps  are scaled so that the $x$- and the $y$-coordinates
 always span  the  interval $[-1, 1]$. The formula used for scaling is
$x_{\rm{sca}}  =2 \cdot (\frac{x - x_{\rm{min}}}{x_{\rm{max}} - x_{\rm{min}}}) - 1$,
 $y_{\rm{sca}}  =2 \cdot (\frac{y - y_{\rm{min}}}{y_{\rm{max}} - y_{\rm{min}}}) - 1$, 
where $x_{\rm{min}}$ and  $x_{\rm{max}}$ are the minimum
 and maximum of  the $x$-coordinates  of all the points in the original map,
 and similarly  for $y_{\rm{min}}$ and $y_{\rm{max}}$.

Each Molecular Distance Map  has some error, that is, 
the Euclidean distances $d_{i, j}$ are not exactly the same as  $f(\Delta(i,j))$.
 When using the same dataset, the error is in general lower for an MDS map in a 
higher-dimensional space.
The {\it Stress-1} (Kruskal stress, \cite{Kruskal64}), is defined in our case as

$$\mbox{{\it Stress-1}} = \sigma_1 = \sqrt{\frac{\Sigma_{i < j} [f(\Delta(i,j)) - d_{i,j}]^2}{\Sigma_{i < j} d_{i,j}^2}}$$

\noindent
where  the summations extend over all the sequences considered for a given map,
 and $f(\Delta(i,j)) = a \times \Delta(i, j) + b$  is a linear 
function whose parameters $a, b \in \mathbb{R}$ are determined by linear 
regression for each dataset and corresponding Molecular Distance Map. 
A benchmark that is often used to assess MDS results  is that  {\it Stress-1}
 should  be in the range  $[0, 0.20]$, see \cite{Kruskal64}.

The dataset  consists of the entire  collection of complete mitochondrial
 DNA sequences from NCBI as of 12 July, 2012.  This
 dataset consists of  3,176 complete mtDNA sequences, namely  79 protists, 111 fungi,
 283 plants, and 2,703 animals. 
This collection of mitochondrial genomes has a great breadth of species  across taxonomic
 categories and great depth of species coverage in certain taxonomic categories.
For example, we compare sequences at every rank of taxonomy, with some pairs being
 different at as high as the (super)kingdom level, and some pairs of
sequences being from the exact same species, as in the case of {\it Silene conica} for
 which our dataset contains the sequences of 140 different mitochondrial chromosomes \cite{Sloan2012}.
 The prokaryotic origins and  
evolutionary history  of mitochondrial genomes have long been extensively studied, 
which will allow comparison of our results
  with both phylogenetic trees and barcodes. Lastly, this genome dataset permits testing
 of both recent and deep rooted species relationships, providing fine resolution of species
 differences.
 
\begin{figure*}[ht]

	\begin{center}
	\includegraphics[width=10cm]{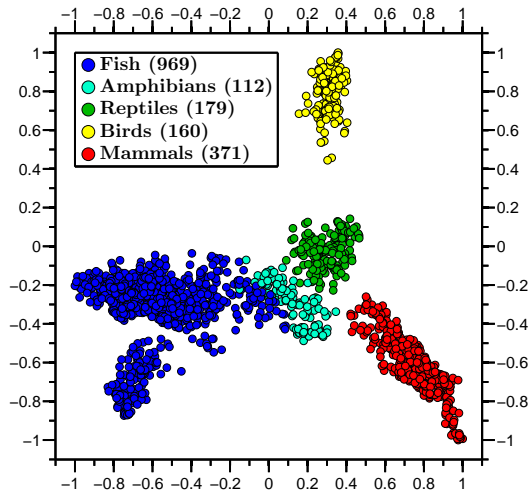} 
		\caption{Molecular Distance Map of 
 phylum Vertebrata (excluding the 5 represented jawless vertebrates), with its
	 	five subphyla.
        The total number of mtDNA sequences is 1,791, the average DSSIM distance is 0.72,
        and  the MDS {\it Stress-1}  is 0.12.}
		\label{vertebrates}
	\end{center}

\end{figure*}

An example of the CGR/DSSIM/MDS approach is the Molecular Distance Map in Figure~\ref{vertebrates}
 which depicts the complete mitochondrial DNA sequences  of all 1,791  jawed vertebrates in our
 dataset. In the legends of Figures 2-7,  the number of represented mtDNA sequences  in each category is listed
 in paranthesis after the category name.
All five different  subphyla of jawed vertebrates  
 are separated in non-overlapping clusters, with very few exceptions.
Examples of fish species mixed with amphibians include {\it Polypterus
ornatipinnis} (\#3125, ornate bichir), {\it Polypterus senegalus} (\#2868, Senegal bichir),
 both  with primitive pairs of lungs; {\it Erpetoichthys calabaricus}
 (\#2745, snakefish) who can breathe atmospheric air using a pair of lungs;
 and {\it Porichtys myriaster} (\#2483, specklefish midshipman) a toadfish of the order
 Batrachoidiformes. It is noteworthy that the question of whether species of the {\it Polypterus} genus are fish or
amphibians has been discussed extensively for hundreds of years \cite{Hall2001}.
 Interestingly, all four represented lungfish (a.k.a. salamanderfish),
 are  mixed with  the amphibians:  {\it Protopterus aethiopicus} (\#873, marbled lungfish),
{\it Lepidosiren paradoxa} (\#2910, South American lungfish),
{\it Neoceratodus forsteri} (\#2957, Australian lungfish),
{\it Protopterus doloi} (\#3119, spotted African lungfish).

The creation of the datasets,  acquisition of data from NCBI's GenBank,  generation of the CGR images,
  calculation of the distance matrix, and  calculation of the Molecular Distance Maps using MDS, 
were all done with the free open-source MATLAB program OpenMDM  \cite{Dattani2013}. 
This program makes use of an open source MATLAB program for SSIM written by Z.Wang \cite{Wang2003},
 and MATLAB's built-in  MDS function$^6$.
\footnotetext[6]{Supplemental Material including the annotated dataset, its DSSIM distance matrix,
 and  full-size versions of Figures 2-7, with numerical mtDNA sequence
identifiers, is available at http://www.csd.uwo.ca/\~{}lila/MapOfLife/}

\section{Results and Discussion}

\begin{figure*}[ht]
 
	\begin{center}
	\includegraphics[width=10cm]{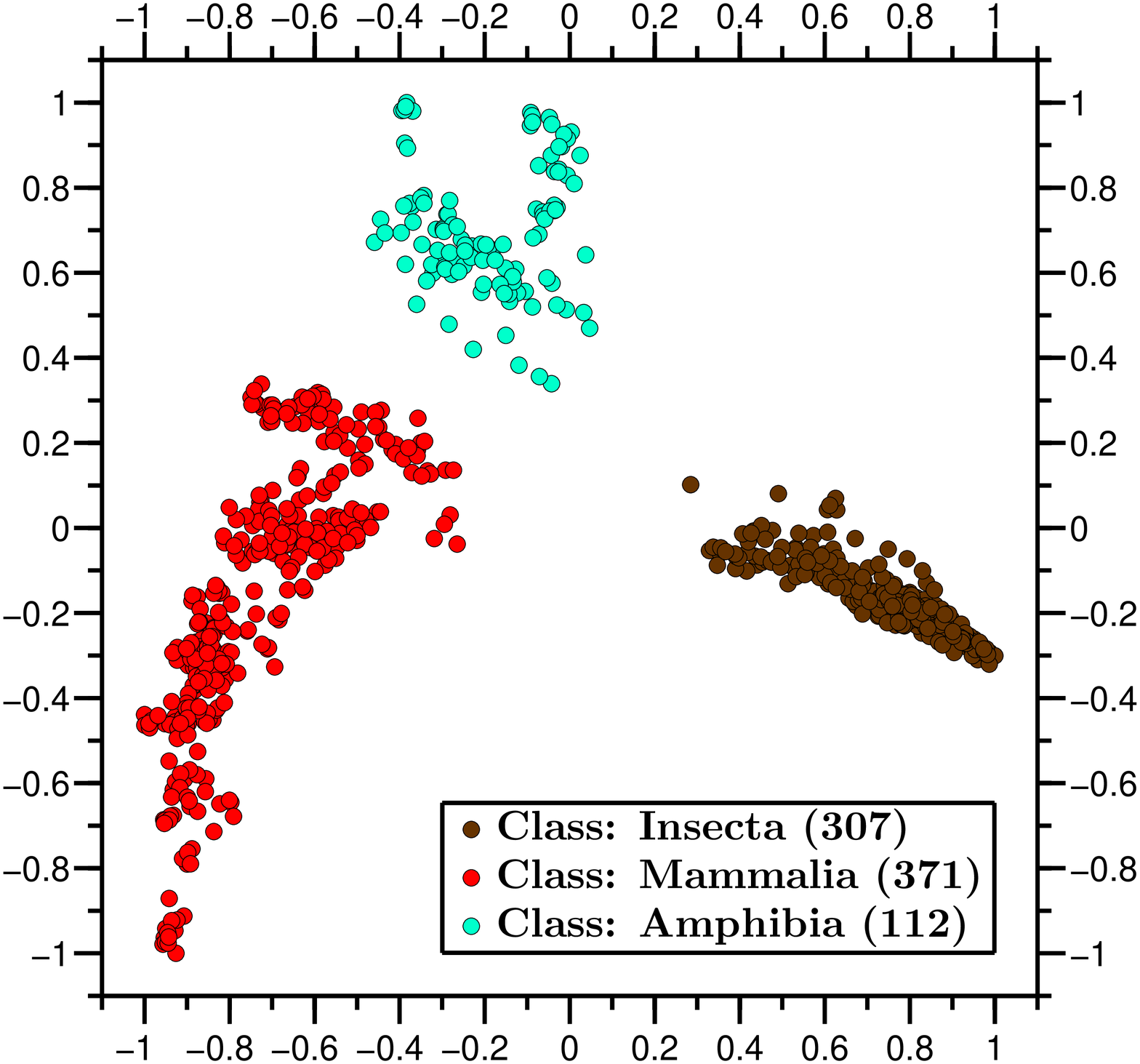} 	
\caption{Molecular Distance Map of three classes: Amphibia, Insecta and Mammalia.
The total number of mtDNA sequences is 790, the average DSSIM distance is 0.65, and the MDS
 {\it Stress-1} is 0.16.}
		\label{insects_mammals_amphibians} 
	\end{center}

\end{figure*}

\begin{figure*}[ht]
	\begin{center}
\includegraphics[width=12cm]{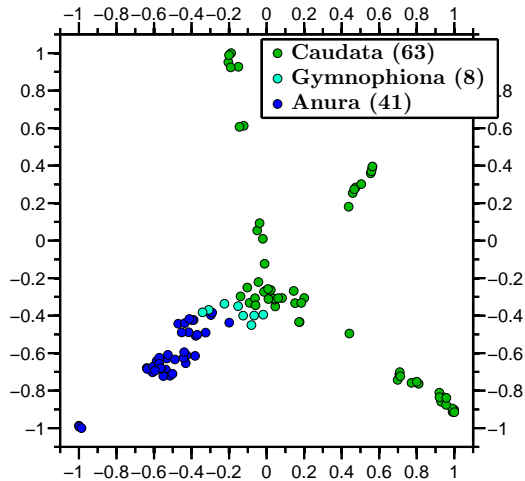} 
		\caption{Molecular Distance Map of Class Amphibia and  three of its  orders.
  The total number of mtDNA sequences is 112,   the average DSSIM distance is  0.70, and the MDS {\it Stress-1} is 0.17.}
	 	\label{amphibians} 
	\end{center}
\end{figure*}

\begin{figure*}[ht]
	\begin{center}
		\includegraphics[width=12cm]{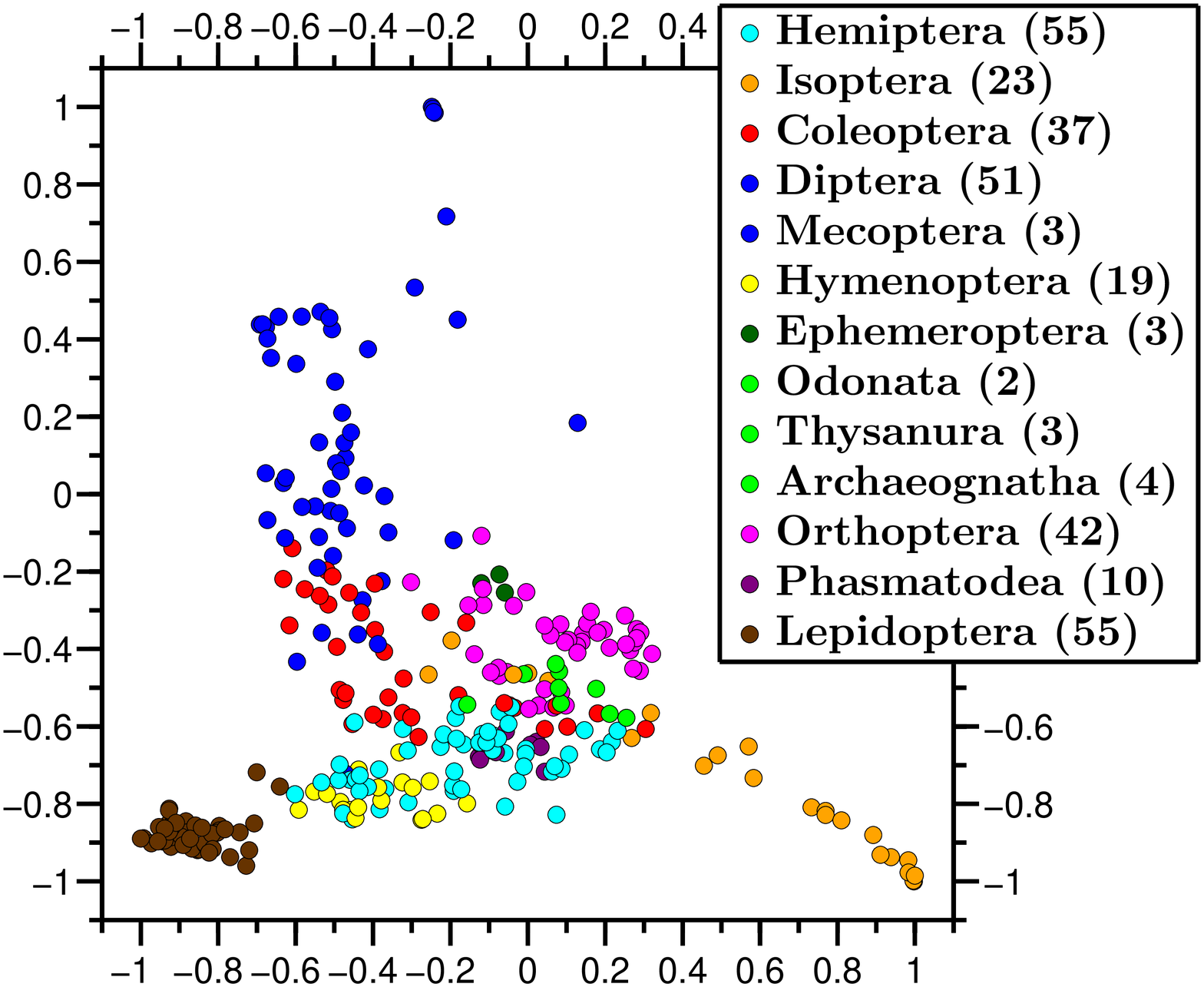} 
                \caption{Molecular Distance Map of Class Insecta.
 The total number of mtDNA sequences is 307,  the average DSSIM distance is 0.52, and the MDS 
  {\it Stress-1} is 0.14.}
		\label{insects}
	\end{center}
\end{figure*}

We applied our method to visualize all available complete mtDNA sequences from
 {\it three classes}, Amphibia, Insecta and Mammalia, in Figure \ref{insects_mammals_amphibians}. 
On a
 finer scale, we applied this method to  observe relationships {\it within a class}:  class 
Amphibia and
  three of its orders in Figure \ref{amphibians}, and class Insecta grouped in  nine categories 
in Figure \ref{insects}.
Note that a feature of  MDS  is that the points $p_{i}$ are not unique.
Indeed, one can translate or rotate a  map without affecting the pairwise Euclidean
distances $d(i, j) = ||p_{i}-p_{j}||$. In addition, the obtained points in an MDS map may change 
coordinates when more data items
 are added to or removed from the dataset. This is because the output of the MDS aims to preserve only the 
pairwise Euclidean distances between points, and this can be achieved even when some
of the points change their coordinates.  In particular, while the position within a taxonomic category may be 
correctly preserved, the  $(x, y)$ coordinates of a  point representing an amphibian species 
in the amphibians-insects-mammals map (Figure \ref{insects_mammals_amphibians}) will not
 necessarily be the same as the $(x, y)$ coordinates of the same  point when amphibians-only
 are mapped (Figure \ref{amphibians}).

\begin{figure*}[ht]
	\begin{center}
\includegraphics[width=12cm]{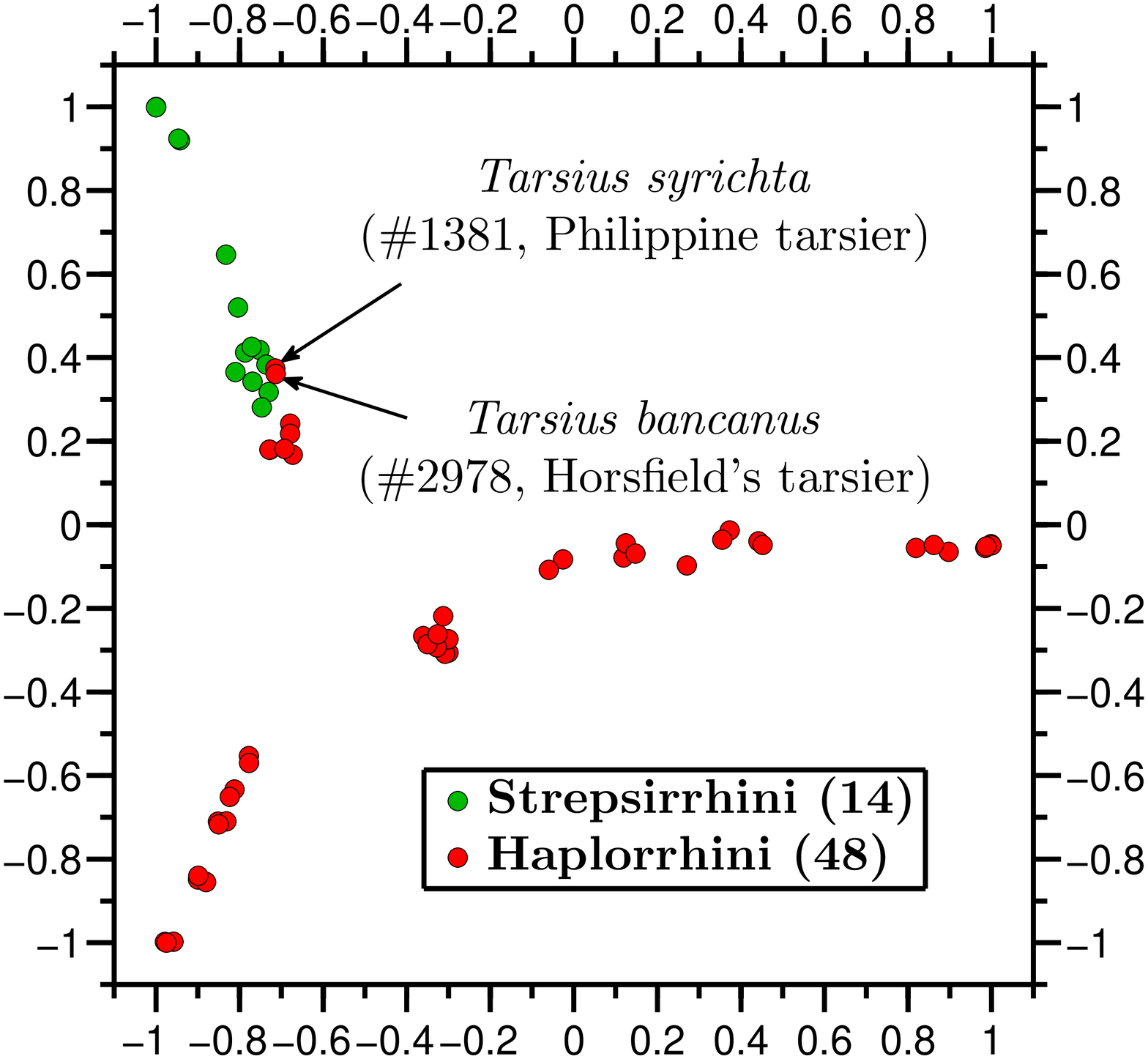}
\caption{Order Primates and its suborders: Haplorhini (anthropoids and tarsiers), and 
 Strepsirrhini (lemurs and lorises). The total number of mtDNA sequences is 62, 
the average DSSIM distance is 0.64,  and the MDS {\it Stress-1} is 0.19.}
		\label{primatesonly}
	\end{center}
\end{figure*}

In general, Molecular Distance Maps are in  good agreement with classical phylogenetic
trees at all scales of taxonomic comparisons,  see  Figure \ref{insects} with \cite{Marshall2006},
  Figure \ref{amphibians} with
 \cite{Pyron2011}, and Figure \ref{primatesonly} with \cite{Shoshani1996}.
Our approach may also provide supporting evidence for a DNA sequence-based classification for
 certain species where morphology-based taxonomy is uncertain.
In addition, our approach may  be able to weight in on conflicts between 
taxonomic classifications  based on morphological traits and those based on more  recent
molecular data, as in the case of tarsiers   (Figure \ref{primatesonly}),
  and protists (Figure  \ref{AllProtists}), as seen below.

Zooming in, we observed the relationships within an order, Primates, with its  suborders
 (Figure \ref{primatesonly}). 
Notably, two extinct species of the genus {\it Homo}  are represented:
{\it Homo sapiens neanderthalensis} and {\it Homo sapiens ssp. Denisova}.
Primates can be classified into two groups,
 Haplorhini  (dry-nosed primates comprising
anthropoids and tarsiers) and Strepsirrhini (wet-nosed primates including lemurs and lorises). 
 The map shows a  clear separation of  these suborders, with the top-left arm of the map
 in Figure \ref{primatesonly},
 comprising the Strepsirrhini. However, there are two  Haplorhini placed
 in the Strepsirrhini cluster, namely  {\it Tarsius bancanus} (\#2978, Horsfield's tarsier) 
 and {\it Tarsius syrichta} (\#1381, Philippine tarsier).  The phylogenetic placement of tarsiers 
within the  order Primates has been  controversial  for over a century, \cite{Jameson2011}.
 According to \cite{Chatterjee2009}, mitochondrial DNA evidence places tarsiiformes 
as a sister group to Strepsirrhini, while in contrast, \cite{Perelman2011} places tarsiers 
 within Happlorhini. In Figure \ref{primatesonly}  the tarsiers are located within the
 Strepsrrhini cluster, which may be due to the fact that they
evolved independently for millions of years, \cite{Shoshani1996}.

Finally, we applied our method to a group whose taxonomic classification has historically 
been challenging: Figure \ref{AllProtists}  depicts  all protist species in the dataset whose
 taxon (as defined in the legend of the figure)
 contained more than one representative. The obvious outlier is {\it Haemoproteus} sp. jb1.JA27 
(\#1466),  sequenced in \cite{Beadell2005} (see also \cite{Valkiunas2010}), and listed as an 
{\it unclassified} organism in the NCBI taxonomy.  Interestingly  this outlier is in the same
 kingdom (Chromalveolata), superphylum (Alveolata), phylum
 (Apicomplexa), and class (Aconoidasida) as the two mtDNA sequences {\it Babesia bovis} T2Bo
  (\#1935), and  {\it Theileria parva}
 (\#3173), that appear grouped with it. It therefore seems plausible that  Molecular Distance
 Maps may shed light on the taxonomical ambiguity of {\it Haemoproteus} sp. jb1.JA27.      

\begin{figure*}[ht]
	\begin{center}
                \includegraphics[width=12cm]{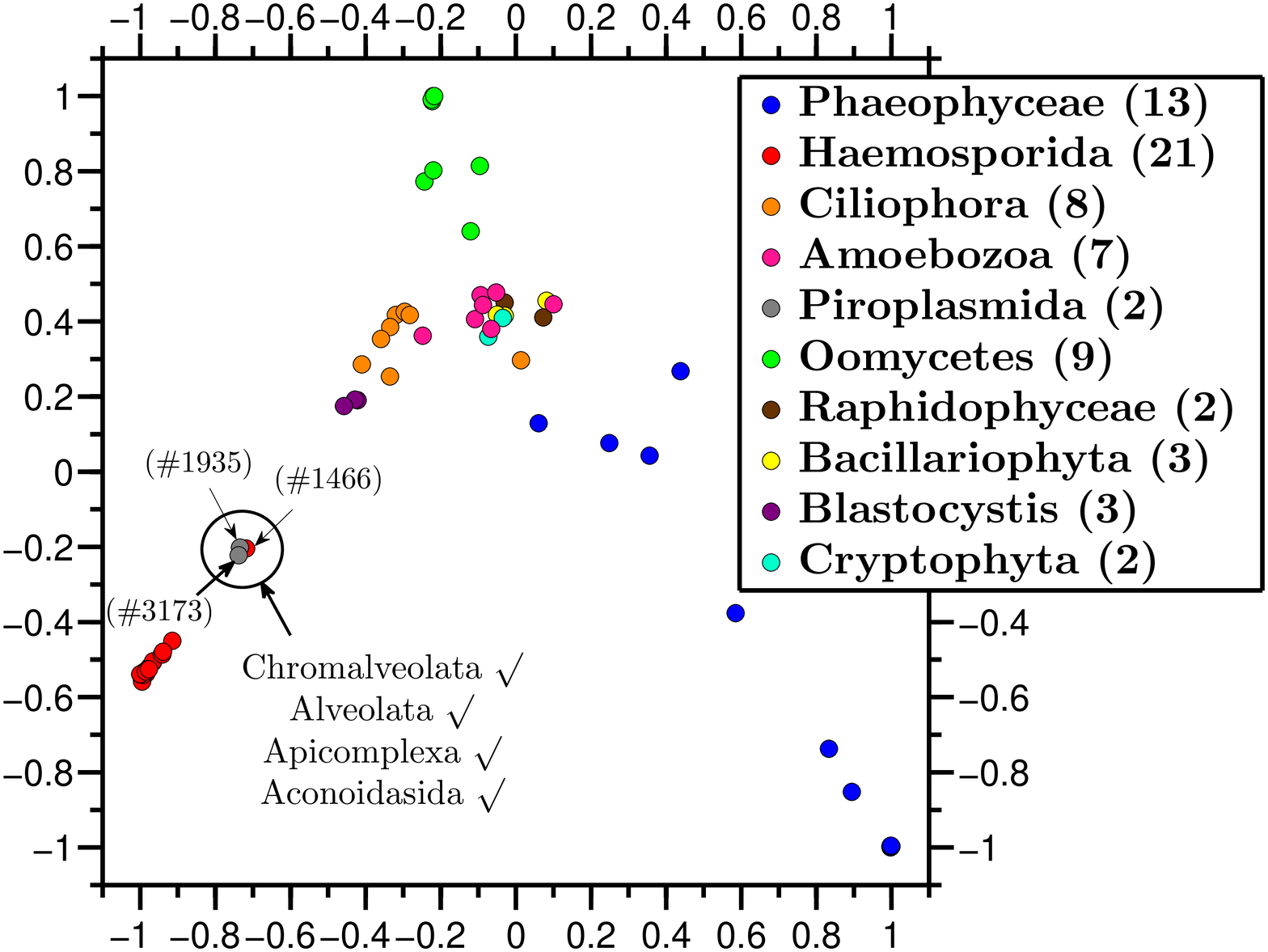} 
		\caption{Molecular Distance Map of all represented species  from  (super)kingdom 
Protista and its orders. The sequence-point \#1466 (red) is  {\it Haemoproteus} sp. jb1.JA27, 
 \#1935 (grey) is {\it Babesia bovis} T2Bo,
and \#3173 (grey) is  {\it Theileria parva}. The annotation shows that all these three 
species belong to the same taxonomic groups, up to the order level. The total number of
 mtDNA sequences is 79, the average DSSIM distance is 0.72,
        and  the MDS {\it Stress-1}  is 0.28.
}
		\label{AllProtists} 
	\end{center}
\end{figure*}

The DSSIM distances computed between all pairs of complete mtDNA sequences varied in range.
The minimum  distance  was 0, between two pairs of identical  mtDNA sequences.
The first pair  comprised the mtDNA of {\it Rhinomugil nasutus} (\#98, shark mullet, length
 16,974 bp)
 and {\it Moolgarda cunnesius} (\#103, longarm mullet,  length 16,974 bp).  A base-to-base sequence
 comparison between these sequences (\#98, NC\_017897.1; \#103, NC\_017902.1) showed  that the sequences
 were indeed identical. However, after completion of this work, the sequence for species \#103 was updated to
 a new version  (NC\_017902.2), on  7 March, 2013, and is now different from the sequence for
 species \#98 (NC\_017897.1). The second pair  comprises the  
 mtDNA  sequences \#1033 and \#1034 (length 16,623 bp), generated by crossing 
 female {\it Megalobrama amblycephala}  with  male {\it Xenocypris davidi}  leading to the creation of
 both diploid (\#1033) and triploid (\#1034) nuclear genomes, \cite{Hu2012}, but identical mitochondrial genomes.

 The maximum distance was found to be
 between   {\it Huperzia squarrosa} (\#118, rock tassel fern, length 413,530 bp) and
  {\it Candida subhashii} (\#954, a yeast,  length 29,795).
Thus, the pair with the  maximum distance $\Delta (\# 118, \# 954) = 0.9969$ featured neither the longest
 mitochondrial sequence,  belonging to {\it Cucumas sativus} (\#533, cucumber, length 1,555,935 bp),
 nor the shortest mitochondrial sequence,  belonging to {\it Silene conica} (\#440, sand catchfly, a plant, 
 length 288 bp).

An inspection  of the   distances between {\it Homo sapiens sapiens} and all the other primate
 mitochondrial genomes in the dataset 
showed that the minimum distance  to {\it Homo sapiens sapiens} was   
 $\Delta (\# 1321, \# 1720) = 0.109$,  the distance to {\it Homo sapiens neanderthalensis} (\#1720,
 Neanderthal),  with the second smallest distance to it being  $\Delta(\# 1321, \# 1052) = 0.18$,
 the distance to {\it Homo sapiens ssp. Denisova} (\#1052, Denisovan). The third
 smallest distance was $\Delta(\# 1321, \# 3084) =  0.4655$ to 
{\it Pan troglodytes} (\#3084, chimp).  Figure 8 shows the graph of the distances between 
the {\it Homo sapiens sapiens} mtDNA and each of the  primate mitochondrial genomes. With no exceptions,
 this graph is in full agreement with established phylogenetic trees, \cite{Shoshani1996}.

\begin{figure*}[ht]
\begin{centering}
\includegraphics[width=15cm]{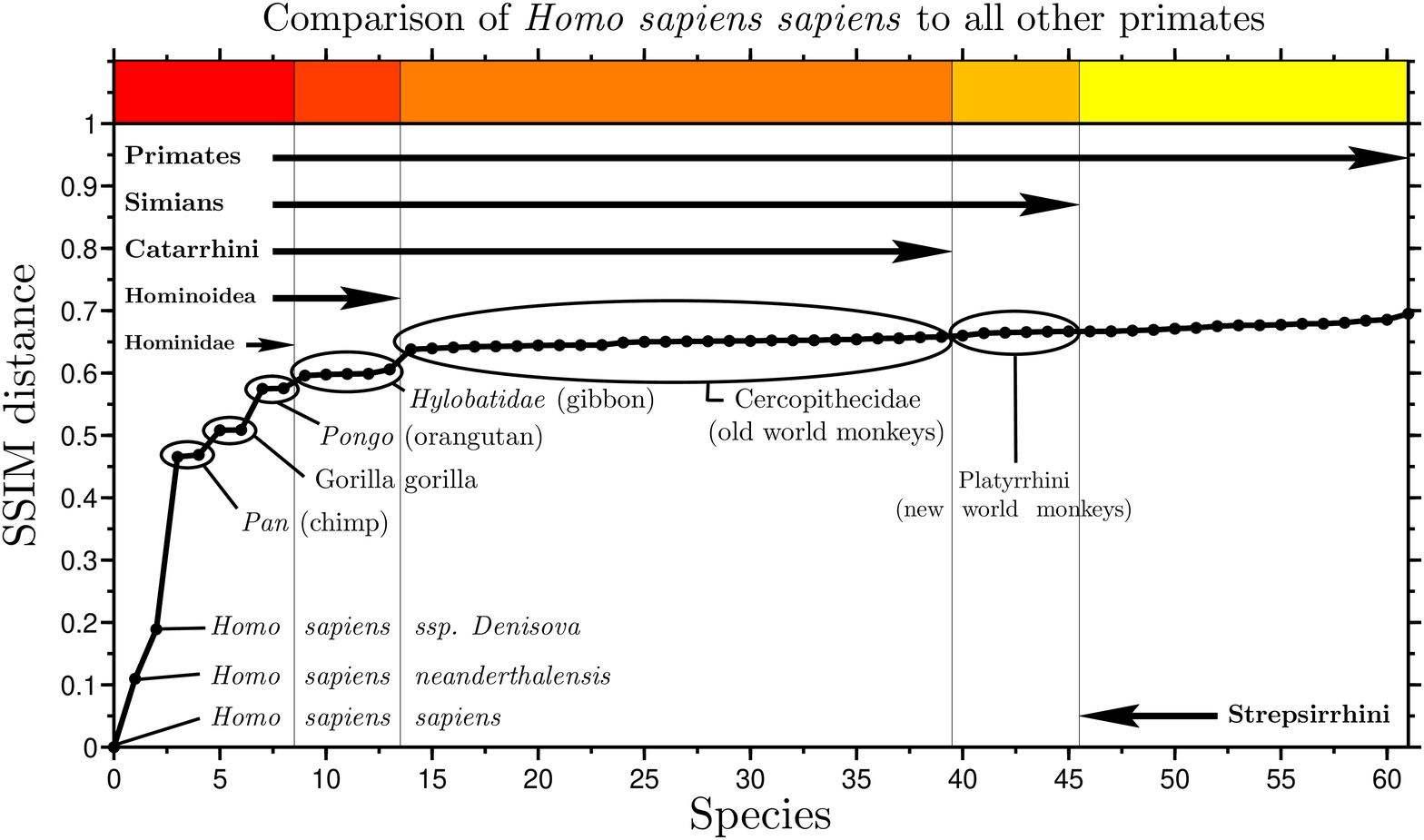} 
\caption{Graph of the DSSIM distances between the CGR images of {\it Homo sapiens sapiens} mtDNA and each of the 62 primate 
 mitochondrial genomes (sorted).} 
\end{centering}
\label{human_graph}
\end{figure*}

In addition to comparing real DNA sequences,  our method can compare  real DNA sequences to
 computer-generated sequences. As an example, we compared the mtDNA genome of 
 {\it Homo sapiens sapiens} with one hundred artificial, computer-generated, DNA sequences of
 the same length and the  same  trinucleotide frequencies as the original. 
The average  distance between these artificial sequences and the original human mitochondrial DNA is 0.9426. 
This indicates that   all  ``human''  artificial DNA sequences are  more distant  from the  {\it Homo sapiens sapiens}
 mitochondrial genome than {\it Drosophila melanogaster } (\#3120, fruit fly) mtDNA, 
with $\Delta (\# 3120, \# 1321) = 0.9313$. This further implies that  trinucleotide frequencies
may not contain sufficient  information to classify  a genetic sequence, suggesting that  
 Goldman's claim \cite{Goldman1993} that ``CGR  gives no futher insight into the structure of the
 DNA sequence than is given by the  dinucleotide and trinucleotide frequencies'' may not hold in general.

 The {\it Stress-1} values for all but one of the  Molecular Distance Maps in this paper  were
  in the ``acceptable'' range $[0, 0.2]$, with the highest value 0.19 in Figure \ref{primatesonly}, where
 the dataset consisted of all 
primates' mtDNA sequences and  the distances were all very small. The exception is Figure \ref{AllProtists} with
{\it Stress-1} equal to 0.28.  Note that   {\it Stress-1} generally decreases with an increase in dimensionality, 
from $q = 2$ to $q = 3, 4, 5...$, and that in   \cite{BorgGroenen2010} it is
 suggested  that it is not always the case that only MDS  representations with {\it Stress-1} under $0.2$ are acceptable,
 nor that all MDS representations with {\it Stress-1}  under $0.05$ are good.

In all the  calculations in this paper, we  used the full mitochondrial sequences.
  However, since  the length of a sequence can influence the brightness of its CGR and
 thus  its Molecular Distance Map coordinates,
 further analysis is needed to elucidate the effect of sequence length on  the positions  of sequence-representing points in a
Molecular Distance Map.  The choice of length of DNA
 sequences used  may ultimately depend  on the particular dataset and particular application.

\section{Conclusions}

Molecular Distance Maps combine three methods (CGR/DSSIM/MDS)  to
measure distances between {\it any} DNA sequences (real or computer-generated) 
of a given dataset, and visually display their interrelationships.
Applications of Molecular Distance Maps  include clarification of taxonomic 
dilemmas, taxonomic classifications,  species identification,
 studies of evolutionary history, as well as possible
 quantitative  definitions of the notions of species and other taxa. 
 
 Possible extensions include generalizations of CGR images to three-dimensional CGR
  images mapped using a regular tetrahedron,
 as well as MDS generalizations to  three-dimensional Molecular Distance Maps  for improved accuracy.
 We  note also that this
 method can be applied to analyzing sequences over other alphabets. For example binary sequences could be imaged
 using a square or a tetrahedron with vertices  labelled 00, 01, 10, 11, and then  DSSIM and MDS could be
 employed to  compare and map them.
Lastly, we note  that the use of the particular image distance measure (DSSIM) or particular scaling technique 
(classical MDS)  does not mean that these are the best choices in all cases, and other image distance measures
 as well as refinements of classical MDS may be explored for optimal results.

\vspace*{3mm}

\noindent
{\small {\it Acknowledgements}.
 We thank Ronghai Tu for an early version of our MATLAB code to generate CGR images, 
 Tao Tao for assistance with NCBI's GenBank,  Steffen Kopecki for generating
 artificial sequences and  discussions. We also thank Andre Lachance, Jeremy McNeill, and Greg Thorn for resources
 and discussions on taxonomy. We thank the Oxford University Mathematical Institute for the use of their Windows compute
 server Pootle/WTS. This work was supported by  Natural Sciences and Engineering Research Council of Canada (NSERC)
 Discovery Grants to L.K. and K.H.; Oxford University Press Clarendon 
Fund and NSERC USRA, PGSM, and PGSD3 awards to N.D.; NSERC USRA award to N.B.}

\bibliography{GenomeMapsCorrect}

\begin{thebibliography}{10}

\bibitem{Beadell2005}
J.~Beadell and R.~Fleischer.
\newblock A restriction enzyme-based assay to distinguish between avian
  hemosporidians.
\newblock {\em Journal of Parasitology}, 91:683--685, 2005.

\bibitem{BorgGroenen2010}
I.~Borg and P.~Groenen.
\newblock {\em {Modern Multidimensional Scaling: Theory and Applications}}.
\newblock Springer, 2nd edition, 2010.

\bibitem{Chatterjee2009}
H.~Chatterjee, S.~Ho, I.~Barnes, and C.~Groves.
\newblock Estimating the phylogeny and divergence times of primates using a
  supermatrix approach.
\newblock {\em BMC Evolutionary Biology}, 9(259), 2009.

\bibitem{Dattani2013}
N.~Dattani, S.~Sayem, R.~Tu, and N.~Bryans.
\newblock Open{MDM}.
\newblock {\em Computer Program}, pages
  https://github.com/ndattani/Dattani--Sayem--Bryans--genomicMDS, 2013.

\bibitem{Deschavanne2000}
P.~Deschavanne, A.~Giron, J.~Vilain, C.~Dufraigne, and B.~Fertil.
\newblock Genomic signature is preserved in short {DNA} fragments.
\newblock In {\em IEEE Intl. Symposium on Bio-Informatics and Biomedical
  Engineering}, pages 161--167, 2000.

\bibitem{Deschavanne1999}
P.~Deschavanne, A.~Giron, J.~Vilain, G.~Fagot, and B.~Fertil.
\newblock Genomic signature: characterization and classification of species
  assessed by {Chaos Game Representation} of sequences.
\newblock {\em Molecular Biology and Evolution}, 16(10):1391--1399, 1999.

\bibitem{Fitch1967}
W.~Fitch and E.~Margoliash.
\newblock {Construction of phylogenetic trees}.
\newblock {\em Science}, 155(760):279--284, 1967.

\bibitem{Gates1986}
M.~Gates.
\newblock A simple way to look at {DNA}.
\newblock {\em J. Theor. Biology}, 119(3):319--328, 1986.

\bibitem{Goldman1993}
N.~Goldman.
\newblock Nucleotide, dinucleotide and trinucleotide frequencies explain
  patterns observed in {Chaos Game Representations} of {DNA} sequences.
\newblock {\em Nucleic Acids Research}, 21(10):2487--2491, 1993.

\bibitem{Hall2001}
B.~Hall.
\newblock {J}ohn {S}amuel {B}udgett (1872-1904): {I}n pursuit of {P}{\it
  olypterus}.
\newblock {\em BioScience}, 51(5):399--407, 2001.

\bibitem{Hebert2003}
P.~Hebert, A.~Cywinska, S.~Ball, and J.~Dewaard.
\newblock Biological identifications through {DNA} barcodes.
\newblock {\em Proc. Biol. Sci}, 270:313--321, 2003.

\bibitem{Hill1992}
K.~Hill, N.~Schisler, and S.~Singh.
\newblock {Chaos Game Representation} of coding regions of human globin genes
  and alcohol dehydrogenase genes of phylogenetically divergent species.
\newblock {\em J. Mol. Evol.}, 35(3):261--9, 1992.

\bibitem{Hill1997}
K.~Hill and S.~Singh.
\newblock Evolution of species-type specificity in the global {DNA} sequence
  organization of mitochondrial genomes.
\newblock {\em Genome}, 40:342--356, 1997.

\bibitem{Hoef2012}
K.~Hoef-Emden.
\newblock Pitfalls of establishing {DNA} barcoding systems in protists: the
  {C}ryptophyceae as a test case.
\newblock {\em PLoS One}, 7:e43652, 2012.

\bibitem{Hollingsworth2009}
P.~Hollingsworth et~al.
\newblock A {DNA} barcode for land plants.
\newblock {\em PNAS}, 106(31):12794--2797, 2009.

\bibitem{Hu2012}
J.~Hu et~al.
\newblock Characteristics of diploid and triploid hybris derived from female
  {M}egalobrama amblycephala {Yih}{X}male {X}enocypris davidi {B}leeker.
\newblock {\em Aquaculture}, 364-365:157--164, 2012.

\bibitem{Jameson2011}
N.~Jameson et~al.
\newblock Genomic data reject the hypothesis of a prosimian primate clade.
\newblock {\em Journal of Human Evolution}, 61:295--305, 2011.

\bibitem{Jeffrey1990}
H.~Jeffrey.
\newblock {Chaos Game Representation of gene structure.}
\newblock {\em Nucleic Acids Research}, 18(8):2163--2170, 1990.

\bibitem{Jeffrey1992}
H.~Jeffrey.
\newblock Chaos game visualization of sequences.
\newblock {\em Comput. Graphics}, 16(1):25--33, 1992.

\bibitem{Kress2005}
W.~Kress, K.~Wurdack, E.~Zimmer, L.~Weigt, and D.~Janzen.
\newblock {Use of {DNA} barcodes to identify flowering plants}.
\newblock {\em PNAS}, 102(23):8369--8374, 2005.

\bibitem{Kruskal64}
J.~Kruskal.
\newblock {Multidimensional scaling by optimizing goodness of fit to a
  nonmetric hypothesis}.
\newblock {\em Psychometrika}, 29(1):1--27, 1964.

\bibitem{PhylogeneticHandbook}
P.~Lemey, M.~Salemi, and A.~Vandamme, editors.
\newblock {\em The Phylogenetic Handbook: A Practical Approach to Phylogenetic
  Analysis and Hypothesis Testing}.
\newblock Cambridge Univ. Press, 2nd edition, 2009.

\bibitem{Leong1995}
P.~Leong and S.~Morgenthaler.
\newblock Random walk and gap plots of {DNA} sequences.
\newblock {\em Computer applications in the biosciences : CABIOS},
  11(5):503--507, 1995.

\bibitem{Lessa1990}
E.~Lessa.
\newblock {Multidimensional analysis of geographic genetic structure}.
\newblock {\em Systematic Zoology}, 39(3):242--252, 1990.

\bibitem{Liao2005}
B.~Liao.
\newblock A {2D} graphical representation of {DNA} sequence.
\newblock {\em Chemical Physics Letters}, 401(1--3):196--199, 2005.

\bibitem{Marshall2006}
S.~Marshall.
\newblock {\em Insects: {T}heir natural history and diversity}.
\newblock Firefly Books, 2006.

\bibitem{ScienceNews12}
S.~Milius.
\newblock New species of the year.
\newblock {\em Science News}, 182(13):30, 2012.

\bibitem{Mora2011}
C.~Mora, D.~Tittensor, S.~Adl, A.~Simpson, and B.~Worm.
\newblock How many species are there on earth and in the ocean?
\newblock {\em PLoS Biology}, 9(8):1--8, 2011.

\bibitem{Nandy1994}
A.~Nandy.
\newblock A new graphical representation and analysis of {DNA} sequence
  structure: Methodology and application to globin genes.
\newblock {\em Current Science}, 66(4):309 -- 314, 1994.

\bibitem{Perelman2011}
P.~Perelman et~al.
\newblock A molecular phylogeny of primates.
\newblock {\em PLoS Genetics}, 7(3), 2011.
\newblock e1001342.

\bibitem{Pyron2011}
R.~Pyron and J.~Wiens.
\newblock A large-scale phylogeny of amphibia including over 2800 species, and
  a revised classification of extant frogs, salamanders, and caecilians.
\newblock {\em Molecular Phylogenetics and Evolution}, 61:543--583, 2011.

\bibitem{Qi2011}
Z.~Qi, L.~Li, and X.~Qi.
\newblock Using {H}uffman coding method to visualize and analyze {DNA}
  sequences.
\newblock {\em Journal of Computational Chemistry}, 32(15):3233--3240, 2011.

\bibitem{Randic2000}
M.~Randic, M.~Vracko, A.~Nandy, and S.~Basak.
\newblock On {3D} graphical representation of {DNA} primary sequences and their
  numerical characterization.
\newblock {\em J. Chem. Inf. and Comp. Sci.}, 40(5):1235--1244, 2000.

\bibitem{Schoch2011}
C.~Schoch et~al.
\newblock Nuclear ribosomal internal transcribed spacer {(ITS)} region as a
  universal {DNA} barcode marker for {F}ungi.
\newblock {\em PNAS}, 109(16):6241--6246, 2012.

\bibitem{Shoshani1996}
J.~Shoshani et~al.
\newblock Primate phylogeny: morphological vs molecular results.
\newblock {\em Molecular Phylogenetics and Evolution}, 5(1):102--154, 1996.

\bibitem{Sirovich2010}
L.~Sirovich, M.~Stoeckle, and Y.~Zhang.
\newblock Structural analysis of biodiversity.
\newblock {\em PLoS ONE}, 5(2):e9266, 2010.

\bibitem{Sloan2012}
D.~Sloan et~al.
\newblock Rapid evolution of enormous, multichromosomal genomes in flowering
  plant mitochondria with exceptionally high mutation rates.
\newblock {\em PLoS Biology}, 10:e1001241, 2012.

\bibitem{Unwin2003}
R.~Unwin and M.~Maiden.
\newblock Multi-locus sequence typing: a tool for global epidemiology.
\newblock {\em Trends Microbiol.}, (11):479--487, 2003.

\bibitem{Valkiunas2010}
G.~Valkiunas et~al.
\newblock A new {H}aemoproteus species ({H}aemosporida: {H}aemoproteidae) from
  the endemic {G}alapagos dove {Z}enaida galapagoensis, with remarks on the
  parasite distribution, vectors, and molecular diagnostics.
\newblock {\em Journal of Parasitology}, 96:783--792, 2010.

\bibitem{Wang2005}
Y.~Wang, K.~Hill, S.~Singh, and L.~Kari.
\newblock The spectrum of genomic signatures: From dinucleotides to {Chaos Game
  Representation}.
\newblock {\em Gene}, 346:173--185, 2005.

\bibitem{Wang2003}
Z.~Wang.
\newblock {SSIM} index.
\newblock {\em Computer Program}, page
  https://ece.uwaterloo.ca/~z70wang/research/ssim/, 2003.

\bibitem{Wang2004}
Z.~Wang, A.~Bovik, H.~Sheikh, and E.~Simoncelli.
\newblock Image quality assessment: From error visibility to structural
  similarity.
\newblock {\em IEEE Transactions on Image Processing}, 13(4):600--612, 2004.

\bibitem{Yao2004}
Y.~Yao and T.~Wang.
\newblock A class of new {2D} graphical representation of {DNA} sequences and
  their application.
\newblock {\em Chemical Physics Letters}, 398(4--6):318--323, 2004.

\bibitem{Yu2010}
C.~Yu, Q.~Liang, C.~Yin, R.~He, and S.~Yau.
\newblock A novel construction of genome space with biological geometry.
\newblock {\em {DNA} Research}, 17(3):155--168, 2010.

\bibitem{Yu2009}
J.~Yu, X.~Sun, and J.~Wang.
\newblock {TN} curve: A novel {3D} graphical representation of {DNA} sequence
  based on trinucleotides and its applications.
\newblock {\em Journal of Theoretical Biology}, 261(3):459 -- 468, 2009.

\bibitem{Yuan2003}
C.~Yuan, B.~Liao, and T.~Wang.
\newblock New {3D} graphical representation of {DNA} sequences and their
  numerical characterization.
\newblock {\em Chemical Physics Letters}, 379:412 -- 417, 2003.

\bibitem{Zhang2012}
Z.~Zhang et~al.
\newblock Colorsquare: A colorful square visualization of {DNA} sequences.
\newblock {\em Comm. in Math. and in Comp. Chemistry}, 68(2):621--637, 2012.

\end{thebibliography}
\bibliographystyle{plain}	

\end{document}